\documentstyle[fleqn,twoside,gc-98,epsfig]{article}

\def\beqa{\begin{eqnarray}}
\def\eeqa{\end{eqnarray}}
\def\beq{\begin{equation}}
\def\eeq{\end{equation}}

\def\half{\frac{1}{2}}

\def\mD{\mbox{\bf D}}
\def\mE{\mbox{\bf E}}
\def\mH{\mbox{\bf H}}
\def\mB{\mbox{\bf B}}
\def\mr{\mbox{\bf r}}

\def\dmunu{_{\mu\nu}}

\def\uab{^{\alpha\beta}}

\def\pa{\partial}

\let\alp=\alpha

\def\pr{{\it Phys. Rev.}\ }
\def\prl{{\it Phys. Rev. Lett.}\ }

\def\cqg{{\it Class. Quantum Grav.}\ }

\def\apj{{\it Ap. J.}\ }
\def\aa{{\it Astron. Astrophys.}\ }

\def\prep{{\it Phys. Rep.}\ }

\def\ie{{\it i.e. }}

\heads{S.Capozziello and G.Iovane}
       {Can quasars be explained by cosmological waveguide effects?}

\begin{document}

\renewcommand{\theequation}{\thesection\arabic{equation}}

\twocolumn[

\Arthead{}{1}{10}

\Title{CAN QUASARS BE EXPLAINED BY \yy
       COSMOLOGICAL WAVEGUIDE EFFECTS?}

\Author{S. Capozziello\foom 1
        and G. Iovane\foom 2}
    {Dipartimento di Scienze Fisiche "E.R. Caianiello",
     Universit\`a di Salerno, \yy
           INFN Sezione di Napoli, Gruppo collegato di Salerno, \yy
           Via S. Allende, I-84081 Baronissi (SA) Italy.}

\Abstract
{A sort of gravitational waveguide effect in cosmology  could explain,
in principle, the huge luminosities
coming from quasars using the cosmological large scale structures
as selfoc--type or planar waveguides. Furthermore,  other
 anomalous phenomena connected with quasars,
 as the existence of
``brothers" or ``twins" objects having different brilliancy
but similar spectra and redshifts, placed on the sky
with large angular distance, could be explained by this effect.
We describe the gravitational waveguide theory and then we
discuss possible realizations in cosmology.}
]
\email 1 {capozziello@vaxsa.csied.unisa.it} \email 2
{geriov@vaxsa.csied.unisa.it}

\section{Introduction}
\sequ{0}

Recently, gravitational lensing has become, in actual fact, a new
field of astronomy and astrophysics to investigate the Galaxy, the
large scale structure of the universe and to test cosmological
models \cite{peebles}. Acting on all scales, it provides a great
amount of applications like a more accurate determination of the
cosmological parameters as $H_{0}$, $\Omega$, and $\Lambda$
\cite{borgeest},\cite{frieman}, the possibility of  describing the
potential of lensing galaxies and galaxy clusters from the
observation of multiply imaged quasars, arcs and arclets
\cite{fort},\cite{broadhurst}. However, the leading role of
gravitational lensing is its contribution in searching for dark
matter. In fact a way to detect compact objects with masses in the
range $10^{-5}M_{\odot}\div 100M_{\odot}$, in the Galaxy or in
nearby galaxies,
 is based upon
an application of lensing, the so called {\it microlensing}, which
effect is to produce characteristic light variations of distant
compact sources. The features of such a curve give the physical
properties of the unseen objects (the Massive Astrophysical
Compact Halo Objects, \ie the MACHOs) which seem greatly to
contribute to the mass of our Galaxy \cite{mollerach}.

Microlensing has also cosmological applications. Particularly
promising are the multiply macro--imaged quasars whose lensing
galaxy should have a large optical depth for lensing effects
\cite{paczynski},\cite{kaiser} \cite{walsh} (at least 20 objects
are identified; see, for example, \cite{blioh},\cite{schneider},
\cite{ehlers}).

The above kinds of analysis are possible if we have a model
explaining the way of forming images such as the above--mentioned
arcs, rings or simply double images and predicting the effects of
the deflector \cite{bartelmann},\cite{kormann}.

From a theoretical point of view, lensing must be treated studying
the geometry of the system source--lens--observer. This study is
simple if we suppose that these are three points on a plane as
well as if we consider  thin  lens approximation: such hypotheses
are reasonable because of the large distances considered. A
theoretical model  can be worked out by giving a specified form to
the lens density, i.e. fixing its structure. From the density
function, using the equations derived from the geometry, we can
have predictions for the observed deviation of the source light
and magnitude of every image.

It is well known that the gravitational lensing may be explained
using the action of a gravitational field on the light rays. In
this case, the action of media with corresponding refraction index
is,  for weak field approximation, completely determined by the
Newtonian gravitational potential which deflects and focuses the
light rays.

In optics, however, there exist other types of devices, like
optical fibers and  waveguides which use the same deflection
phenomena. The analogy with the action of a gravitational field
onto light rays may be extended to incorporate these other
structures on the light. In other words, it is
 possible to suppose the existence of a sort of
gravitational waveguide effect \cite{dodma},\cite{dodonov},
\cite{dodma1}. furthermore, structures like cosmic strings,
texture and domain walls,
 which are produced at phase transiton in inflationary
models, can evolve into today observed filaments, clusters and
groups of galaxies and behave in a variety of ways with respect to
the propagation of light. In fact, the lensing by cosmic string
was suggested as explaination of the observation \cite{turner} of
twins objects with very large angular distance between the
partners \cite{vilenkin},\cite{gott}.

The aim of this work is to discuss the properties of possible
waveguides in the universe and to suggest the explaination of some
phenomena, like quasar huge luminosities and large angular
distances between twins, as a by--product of their existence. For
example, a filament of galaxies can be considered  a sort of
waveguide preserving total luminosity of a source, if we have,
locally, an effective gravitational potential of the form
$\Phi(\bf{r})\sim r^{2}$, while the planar structures generated by
the motion of cosmic strings (the so called "wakes") can yield
cosmological structures where the total flux of light is preserved
and the brightness of objects at high redshift, whose radiation
passes through such structures, appears higher to a far observer.

Sec. 2 is devoted to the discussion of the gravitational potential
intended as the refraction index of geometrical optics. In Sec. 3,
we derive the Helmholtz scalar equation, starting from the Maxwell
equation for the electromagnetic field, which is the dynamical
equation for the optical waveguides.

In Sec. 4, we costruct the optical waveguide model using the
paraxial approximation (the so called Fock--Leontovich
approximation \cite{fock}). We propose a model in which we
consider, instead of a simple gravitational lensing effect, the
effect of a sort of system of lenses which, combined in files or
in planes, results as a waveguide. In Sec. 5, we discuss the
eventual cosmological realization of such structures and the
connection with observations, in particular with quasars, giving
some simulations. Conclusions are drawn in Sec.5.

\section{The propagation of light in a weak gravitational field}
\sequ{0}

The behaviour of the electromagnetic field without sources
in the presence of a gravitational field is described by the
Maxwell equations \cite{skrotsky},\cite{plebansky} \beq \label{14}
\frac{\pa F_{\alp\beta}}{\pa x^{\gamma}}+ \frac{\pa
F_{\beta\gamma}}{\pa x^{\alp}}+ \frac{\pa F_{\gamma\alpha}}{\pa
x^{\beta}}=0\;; \eeq \beq \label{15} \frac{1}{\sqrt{-g}}
\frac{\pa}{\pa x^{\beta}}\left(\sqrt{-g}
F^{\alpha\beta}\right)=0\,, \eeq where $F\uab$ is the
electromagnetic field tensor and $\sqrt{-g}$ is the determinant of
the four--dimensional metric tensor. For a static gravitational
field, these equations can be reduced to the usual Maxwell
equations describing the electromagnetic field in media where the
dielectric and magnetic tensor permeabilities are connected with
the metric tensor $g\dmunu$ by the equation \cite{landau}
\beq
\varepsilon_{ik}=\mu_{ik}=-g_{00}^{-1/2}[\mbox{det}g_{ik}]^{-1/2}g_{ik}\;;
\label{16}
\eeq
\[
i,k=1,2,3\,.
\]
If one has an isotropic model, the
metric tensor is diagonal and the refraction index  may be
introduced by mimicking  the gravitational field \beq \label{17}
n({\bf r})=(\varepsilon\mu)^{1/2}\,, \eeq (it is worthwhile to
note that such a situation can be easily reproduced in cosmology
\cite{ehlers}).

For weak gravitational fields, considered also to describe usual
gravitational lensing effects, the metric tensor components are
expressed in terms of the Newton gravitational potential $\Phi$ as
\cite{blioh},\cite{ehlers},\cite{landau}
\beq
\label{19}
g_{00}\simeq 1+2\frac{ \Phi ({\bf r})}{c^{2}}\;;
\eeq
\beq
\label{191}
g_{ik}\simeq -\delta_{ik}\left(1-2\frac{\Phi({\bf
r})}{c^{2}}\right)\;;
\eeq
where we are assuming the weak field
$\Phi/c^{2}\ll 1$ and the slow motion approximation $|v|\ll c$.
Then, due to relations (\ref{16}), (\ref{19}) and (\ref{191}), the refraction
index $n({\bf r})$ in (\ref{17}) can be expressed in terms of the
gravitational potential $\Phi({\bf r})$ produced by some matter
distribution. Such a weak field situation is realized for
cosmological structures which give rise to the gravitational
lensing effects connected to several observable phenomena
(multiple images, magnification, image distorsion, arcs and
arclets) \cite{ehlers}. Here, we are interested to a specific
application which could be realized by some kinds of gravitational
systems as cosmological string--like or planar--like distributions
of matter.

\section{The Helmholtz Equation}
\sequ{0}

In this section, we derive the Helmholtz equation from the Maxwell
equations in media reducing the equation with interaction of
different light polarizations to a scalar equation. Let us write
the Maxwell equations for the electromagnetic field in media
without sources, \ie $\mbox{\bf J}=0, \rho=0$, have the form \beqa
\mbox{rot {\bf E}({\bf r},t)}&=&-\frac{1}{c} \frac{\pa \mbox{\bf
B}({\bf r},t)}{\pa t}\,,\label{m1}\\ \mbox{rot {\bf H}({\bf
r},t)}&=&\frac{1}{c} \frac{\pa \mbox{\bf D}({\bf r},t)}{\pa
t}\,,\label{m2}\\ \mbox{div {\bf B}({\bf r},t)}&=&0\,,\label{m3}\\
\mbox{div {\bf D}({\bf r},t)}&=&0\,,\label{m4} \eeqa where the
media contribution is taken into account by the relations
\beq
\label{m51}
\mD_{\omega}(\mr)=\varepsilon
(\omega,\mr)\mE_{\omega}(\mr)\,,
\eeq
\beq
\label{m5}
\mB_{\omega}(\mr)=\mu
(\omega,\mr)\mH_{\omega}(\mr)\,, \eeq
where the subscript $\omega$
means the Fourier amplitudes of the fields, \ie \beq \label{m6}
\mD(\mr,t)=\int \mD_{\omega}(\mr)e^{-i\omega t}d\omega\,, \eeq and
analogously for $\mE, \mB, \mH$. Then, taking the Fourier
transforms with respect to the time variable, we get \beqa
\mbox{rot} \mE_{\omega}(\mr)&=&\frac{i\omega}{c}
\mB_{\omega}(\mr)\,,\label{m7}\\ \mbox{rot}
\mH_{\omega}(\mr)&=&-\frac{i\omega}{c}
\mD_{\omega}(\mr)\,,\label{m8}\\
\mbox{div}\mB_{\omega}(\mr)&=&0\,,\label{m9}\\ \mbox{div}
\mD_{\omega}(\mr)&=&0\,.\label{m10} \eeqa
Using the relations
(\ref{m51}), (\ref{5}) we get \beqa
\mbox{rot}\mE_{\omega}(\mr)&=&\frac{i\omega}{c}
 \mu(\omega,\mr)\mH_{\omega}(\mr)\,,\label{m11}\\
\mbox{rot}\mH_{\omega}(\mr)&=&-\frac{i\omega}{c}
\varepsilon(\omega,\mr)\mE_{\omega}(\mr)\,,\label{m12}\\
\mu\,\mbox{div}\mH_{\omega}(\mr)&=&0\,,\label{m13}\\
\mbox{div}[\varepsilon(\omega,\mr)\mE_{\omega}(\mr)]&=&0\,.
\label{m14} \eeqa Let us consider the magnetic permeability $\mu$
to be constant.
Being the operator equality
\beq
\mbox{rot}\,\mbox{rot}=\mbox{grad}\,\mbox{div}-\triangle,
\eeq
we
get from Eq.(\ref{m11})
\[
\mbox{grad}\,\mbox{div}\mE_{\omega}(\mr)-\triangle\mE_{\omega}(\mr)=
\]
\beq
\label{m15}
=\frac{\omega^{2}}{c^{2}}\mu(\omega,\mr)\varepsilon(\omega,\mr)
\mE_{\omega}(\mr)\,.
\eeq
Introducing the refractive index
\beq
n^2(\omega,\mr)=\mu(\omega,\mr)\varepsilon(\omega,\mr)\,,
\eeq we
can rewrite Eq.(\ref{m15}) as
\[
\triangle
\mE_{\omega}(\mr)+\frac{\omega^2}{c^2}n^{2}(\omega,\mr)
\mE_{\omega}(\mr)=
\]
\beq
\label{m16}
=- \nabla\left[\frac{\mE_{\omega}(\mr)
\nabla\varepsilon(\omega,\mr)}{\varepsilon(\omega,\mr)}\right]\,.
\eeq
Here we have used the relation \beq \label{m17}
\mbox{div}\mE_{\omega}(\mr)=-\frac{\mE_{\omega}(\mr)
\nabla\varepsilon(\omega,\mr)}{\varepsilon(\omega,\mr)}\,. \eeq
One can neglect the term in the right--hand side of Eq.(\ref{m16})
if it is much less than both terms in the left--hand side of the
same relation. In fact, since $\triangle=\nabla\cdot\nabla$ for
distances of an order of the light wavelength $\lambda$, the both
terms in the left--hand side of Eq.(\ref{m16}) (independently of
the light polarization) are of the order
\beq
\label{m181}
|\triangle\mE_{\omega}(\mr)|\sim\lambda^{-2}E_{\omega}(\mr)\,,
\eeq
\beq
\label{m18}
\frac{\omega^{2}}{c^2}n^2(\omega,\mr)\mE_{\omega}(\mr)
\sim\lambda^{-2}E_{\omega}(\mr)\,.
\eeq
The term depending on the
light polarization interaction for the same distances is of the
order \beq \label{m19} \nabla\left[\frac{\mE_{\omega}(\mr)
\nabla\varepsilon(\omega,\mr)}{\varepsilon(\omega,\mr)}\right]\sim
\lambda^{-2}\frac{\delta\varepsilon}{\varepsilon}E_{\omega}(\mr)\,,
\eeq where $\delta\epsilon$ is the change of the dielectric
permeability for distances of the order of wavelength $\lambda$.

Comparing Eqs.(\ref{m181}),(\ref{m18}) and (\ref{m19}), we conclude that for
\beq \label{m20} \frac{\delta \varepsilon}{\varepsilon}\ll 1\,,
\eeq we can neglect the term depending on the light polarization
interaction with respect to the other two terms. In this
approximation we get the scalar Helmholtz equation for all the
decoupled components of the electric vector field, \ie \beq
\label{m21} \triangle
\mE_{\omega}(\mr)+\frac{\omega^2}{c^2}n^2(\omega,\mr)
\mE_{\omega}(\mr)=0\,. \eeq If one has a solution of the Helmholtz
equation $\mE_{\omega}^{(0)}(\mr)$, either exact or approximate
one, the influence of the light polarization interaction may be
taken into account the Born method of iteration. In fact, the
Green function given by Eq.(\ref{m21}) $G(\mr,\mr',\omega)$ or by
an approximation of this equation, satisfies the equation \beq
\label{m22}
\left[\triangle+\frac{\omega^{2}}{c^2}n^2(\omega,\mr)\right]G(\mr,\mr',\omega)=
\delta(\mr-\mr')\,. \eeq Then the solution of the equation with
the polarization term has the form
\[
\mE_{\omega}(\mr)=\mE_{\omega}^{(0)}(\mr)+
\]
\beq
\label{m23}
+\int G(\mr,\mr',\omega)
\nabla\left[\frac{\mE_{\omega}^{(0)}(\mr)
\nabla\varepsilon(\omega,\mr)}{\varepsilon(\omega,\mr)}\right]d\mr.
\eeq
Eq.(\ref{m22}) has the form equivalent to the equation for
the Green function of the Schr\"odinger equation for the energy
constant $E$ equal to zero. In fact, if we write down the
Hamiltonian operator \beq \label{m24} \hat{\cal
H}=-\half\triangle+U(\mr)\,, \eeq with $\hbar=m=1$, and the
equation for the Green function of the Schr\"odinger equation
$G_{s}(\mr,\mr',E)$ which is the matrix element of the operator
$(\hat{\cal H}-E)^{-1}$ in the coordinate representation \beq
\label{m25} G_{s}(\mr,\mr',E)=\langle\mr|(\hat{\cal
H}-E)^{-1}|\mr'\rangle\,, \eeq which comes from the equation \beq
\label{m26} (\hat{\cal H}-E)G_{s}(\mr,\mr',E)=\delta(\mr-\mr')\,.
\eeq The comparison of this equation in explicit form \beq
\label{m27}
\left\{-\half\triangle+U(\mr)-E\right\}G_{s}(\mr,\mr',E)=\delta(\mr-\mr')
\eeq
with Eq.(\ref{m22}) shows that they are identical for $E=0$
with the replacements
\beq
\label{m281}
U(\mr)=-2\frac{\omega^2}{c^2}n^2(\omega,\mr)\,,
\eeq
\beq
\label{m28}
-2G(\mr,\mr')=G_{s}(\mr,\mr',0)\,.
\eeq
Thus, we have shown that
if one knows the Green function $G_{s}(\mr,\mr',E)$ of the
Schr\"odinger equation for the unit mass particle moving in a
potential like that in Eq.(\ref{m281}), the Green function of the
Helmholtz equation (\ref{m22}) is given by the equality \beq
\label{m29} G(\mr,\mr')=-\frac{1}{2}G_{s}(\mr,\mr',E=0)\,. \eeq
Since the Green function for the Schr\"odinger equation are
studied for many potentials, the results obtained in quantum
mechanics can be applied for our purposes to study polarization
and waveguiding effects since they are formally identical.

\section{The gravitational waveguide model}
\sequ{0}

Following the above procedure for deriving the scalar Helmholtz
equation  for the components of the electromagntetic field from
the first order Maxwell equations, we get (for some arbitrary
monochromatic component of the electric field) \beq \label{18}
\frac{\pa^{2}E}{\pa z^{2}}+\frac{\pa^{2}E}{\pa x^{2}}
+\frac{\pa^{2}E}{\pa y^{2}}+k^{2}n^{2}({\bf r})E=0\,, \eeq where
$k$ is the wave number.
 This procedure works if, as we have seen,  the relative change of
diffraction index on distances of the light wavelength  is small.

The coordinate $z$, in Eq.(\ref{18}), is considered as the
longitudinal one and it can measure the space distance along the
gravitational field structure produced by a mass distribution with
an optical axis. Such a coordinate may also correspond to a
distance along the light path inside a planar gravitational field
structure produced by a planar matter--energy distribution in some
regions of the universe. In other words, if one has a matter
distribution with some axis like a cylinder with  dust or like a
planar slab with dust, it is possible to consider the
electromagnetic field radiation propagating paraxially. The
parabolic approximation \cite{fock} is used for describing light
propagation in media and in devices as optical fibers
\cite{lifsits}. Below, we will discuss the possibility to use this
approximation for describing electromagnetic radiation propagating
in a weak gravitational field.

Let us consider, the scalar equation (\ref{18}) and the electric
field $E$ of the form
\beq
\label{1}
E=n_{0}^{-1/2}\Psi\exp\left(ik\int^{z}n_{0}(z')dz'\right)\;;
\eeq
\[
n_{0}\equiv n(0,0,z)\,,
\]
where $\Psi(x,y,z)$ is a slowly varying spatial amplitude
along the $z$ axis, and $exp(iknz)$ is a rapidly oscillating phase
factor. Its clear that the beam propagation  is along the $z$
axis. We rewrite Eq.(\ref{18})  neglecting second order derivative
in longitudinal coordinate $z$ and obtain a Schr\"odinger--like
equation for $\Psi$:
\beq
\label{2}
i\lambda\frac{\pa \Psi}{\pa\xi}=
\eeq
\[
=-\frac{\lambda^{2}}{2} \left(\frac{\pa^{2}\Psi}{\pa
x^{2}}+\frac{\pa^{2}\Psi}{\pa y^{2}}\right)
+\frac{1}{2}\left[n_{0}^{2}(z)-n^{2}(x,y,z)\right]\Psi,
\]
where $\lambda$ is the electromagnetic radiation wavelength and we
adopt the new variable \beq \label{3}
\xi=\int^{z}\frac{dz'}{n_{0}(z')}\,, \eeq normalized with respect
to the refraction index \cite{dodma} (for our application,
$n_{0}(z)\simeq 1$ so that $\xi$ coincides essentially with $z$).

At this point, it is worthwhile to note that if one has the
distribution of the matter in the form of cylinder with a constant
(dust) density $~\rho_{0} ~$,  the gravitational potential inside
 has a  parabolic  profile providing waveguide effect
for electromagnetic radiation analogous to sel--foc optical
waveguides realized in fiber optics. In this case,
Schr\"odinger--like equation is that of two--dimensional quantum
harmonic oscillator for which the mode solutions exist in the form
of Gauss--Hermite polynomials (see, for example, \cite{manko}). In
the case of inhomogeneous longitudinal dust distribution in the
cylinder (that is $~\rho \,(z)\,$),
 the
Schr\"odinger-like equation describes the model of two-dimensional
parametric oscillator for which the mode solutions, in the form of
modified Gaussian and Gauss--Hermite polynomials, exist with
parameters determined by the density dependence on longitudinal
coordinate.

As a side remark, it is interesting to stress that, considering
again Eq.(\ref{2}), the term in square brackets in the rhs plays
the role of the potential in a usual Schr\"odinger equation; the
role of Planck constant is now assumed by $\lambda$. Since the
refraction index can be expressed in terms of the Newtonian
potential when we consider the propagation of light in a
gravitational field, we can write the  potential in (\ref{2}) as
\beq \label{pot} U({\bf
r})=\frac{2}{c^{2}}[\Phi(x,y,z)-\Phi(0,0,z)]\,. \eeq The waveguide
effect depends specifically on the shape of potential (\ref{pot}):
for example,  the radiation from a remote source does not
attenuate if $U\sim r^{2}$; this situation is realized supposing a
"filamentary" or a "planar" mass distribution with constant
density $\rho$. Due to the Poisson equation, the potential inside
the filament is a quadratic function of the transverse
coordinates, that is of $r=\sqrt{x^{2}+y^{2}}$ in the case of the
filament and of $r=x$ in the case of the planar structure
(obviously the light propagates in the "remaining" coordinates:
$z$ for the filament, $z,y$ for the plane). In other words, if the
radiation, travelling from some source, undergoes a waveguide
effect, it does not attenuate like $1/R^{2}$ as usual, but it is,
in some sense conserved; this fact means that the source
brightness will turn out to be much stronger than the brightness
of analogous objects located at the same distance (i.e. at the
same redshift $Z$) and the apparent energy released by the source
will be anomalously large.

To fix the ideas, let us estimate how the  electric field
(\ref{1}) propagates into an ideal  filament whose internal
potential is \beq \label{internal}
U(r)=\frac{1}{2}\omega^{2}r^{2}\,,\;\;\;\;\;\;
\omega^{2}=\frac{4\pi G \rho}{c^{2}} \eeq where $\rho$ is constant
and $G$ is the Newton constant.
 A spherical wave from a source,
\beq \label{sphere} E=(1/R)\exp(ikR)\,, \eeq can be represented in
the paraxial approximation as \beq \label{4}
E(z,r)=\frac{1}{z}\exp\left(ikz+\frac{ikr^{2}}{2z}-
\frac{r^{2}}{2z^{2}}\right)\,, \eeq where we are using the
expansion
\beq \label{5} R=\left(z^{2}+r^{2}\right)^{1/2}\approx
z\left(1+\frac{r^{2}}{2z^{2}}\right),\;\;r\ll z\,. \eeq
It
is realistic to  assume $n_{0}\simeq 1$ so that, from (\ref{3}),
$\xi=z$. Assume now that the starting point of the filament of
length $L$ is at a distance $l$ from a source shifted by a
distance $a$ from the filament axis in the $x$ direction. The
amplitude  $\Psi$ of the field $E$, entering the wave guide is
\beq \label{6} \Psi_{in}=\frac{1}{l}\exp
\left[\frac{ikl-1}{2l^{2}}\left((x-a)^{2}+y^{2}\right)\right]\,,
\eeq and so in (\ref{sphere}), we have $
R=\left(l^{2}+y^{2}+(x-a)^{2}\right)^{1/2}.$

We can calculate the amplitude of the field at the exit of the
filament by the equation
\beq
\Psi_{f}(x,y,l+L)=
\label{8}
\eeq
\[
=\int
dx_{1}dy_{1}G(x,y,l+L,x_{1},y_{1},l)\Psi_{in}(x_{1},y_{1},l),
\]
where $G$ is the Green function of Eq.(\ref{2}). For the
potential (\ref{internal}), $G$ has the form
\beq
\label{9}
G(x,y,l+L,x_{1},y_{1},l)
=\frac{\omega}{2\pi
i\lambda\sin\omega\lambda}\times
\eeq
\[
 \exp\left(\frac{i\omega[\cos\omega L
(x^{2}+y^{2}+x_{1}^{2}+y_{1}^{2})-2(xx_{1}+yy_{1})]}{2 \pi
i\lambda\sin\omega\lambda}\right),
\]
which is the propagator
of the harmonic oscillator. The integral (\ref{8}) is Gaussian and
can be exactly evaluated
\beq
\Psi_{f}=\frac{\omega
l}{\omega l^{2}\cos\omega L+(l+i\lambda)\sin\omega l}\times
\label{10}
\eeq
\[
\times\exp\left(-\frac{(x^{2}+y^{2}) [(\omega l
k)^{2}-\omega k(i+k l)\cot\omega l]}{2(1-ikl-ik\omega l^{2}
\cot\omega l)}\right)
\]
\[
\times\exp\left(\frac{a^{2}\omega k(i+k l)\cot\omega
L}{2(1-ikl-ik\omega l^{2}\cot\omega l)}\right)
\]
\[
\times \exp \left(-\frac{2xa\omega k(1+kl)}{2\sin\omega
L(1-ikl-ik\omega l^{2}\cot\omega L)}\right)\,.
\]
The parameter
$l$ drops out of the denominator of the pre--exponential factor if
the length $L$ satisfies the condition \beq \label{11} \tan\omega
L=-\omega l\;. \eeq Eq.(\ref{10}) is interesting in two limits. If
$\omega l\ll 1$, we have
\beq \label{112}
\Psi_{f}=\frac{1}{i\lambda}\exp
\left\{-\frac{l+i\lambda}{2\lambda^{2}l}\left[(x+a)^{2}+y^{2}\right]\right\}\,,
\eeq
which means that the radiation emerging from a point with
coordinate $(a,0,0)$ is focused near  a point with coordinates
$(-a,0,l+L)$ (that is the radius has to be of the order of the
wavelength). This means that, when the beam from an extended
source is focused inside the waveguide in such a way that, at a
distance $L$, Eq.(\ref{11}) is satisfied,  an inverted image of
the source is formed,
 having the very
same geometrical dimensions of the source. The waveguide "draws"
the source closer to the observer since, if the true distance of
the observer from the source is $R$, its image brightness will
correspond to that of a similar source at the closer distance
\beq
\label{eff} R_{eff}=R-l-L\,. \eeq
If we do not have $\omega l\ll
1$, we get (neglecting the term $i\lambda/l$ compared with unity)
\beq
\Psi_{f}=\frac{\sqrt{1+(\omega
l)^{2}}}{i\lambda}\times \label{13}
\eeq
\[
\times\exp \left\{-\frac{1+(\omega
l)^{2}}{2\lambda^{2}} \left[y^{2}+\left(x+\frac{a}{\sqrt{1+(\omega
l)^{2}}}\right)^{2} \right]\right\},
\]
from which, in
general, the size of the image is decreased by a factor
$\sqrt{1+(\omega l)^{1/2}}$. The amplitude increases by the same
factor, so that the brightness is $(R/R_{eff})$ times larger.

In the opposite limit $\omega l\gg 1$, we have $\tan\omega
L\rightarrow\infty$, so that $L\simeq \pi/\omega$, that is the
shortest focal length of the waveguide is \beq \label{foc}
L_{foc}=\sqrt{\frac{\pi c^{2}}{4G\rho}}\,, \eeq which is the
length of focusing of the initial beam of light trapped by the
gravitational waveguide. All this arguments apply if the waveguide
has (at least roughly) a cylindrical geometry. The theory of
planar waveguide is similar but we have to consider only $x$ as
transverse dimension and not also $y$.
The cosmological feasibility of a waveguide depends on the
geometrical dimensions of the structures, on the connected
densities and on the limits of applicability of the above
idealized scheme. In the next section, we shall discuss these
features and the possible candidates which could give rise to
observable effects.

\section{Cosmic structures as waveguides and quasars}
\sequ{0}

The gravitational waveguide effect has the same physical reason
that has the gravitational lens effect which is the
electromagnetic wave deflection by the gravitational field
(equivalent to the deflection of light by  refractive media).
However, there are essential differences producing specific
predictions for observing the waveguide effect. The gravitational
lenses are usually considered as compact objects with strong
enough gravitational potential. The light rays deflected by
gravitational lenses move outside of the matter which forms the
gravitational lens itself. The gravitational waveguide as well as
optical waveguide is noncompact long structure which may contain
small matter density and the deflection of light by each element
of the structure is very small. Due to very large scale sizes of
the structure (we give an extimation below),
 the electromagnetic
radiation deflection by the gravitational waveguide occurs and, in
principle, it may be observed. We will mention, for example, a
possibility of brilliancy magnification of the long distanced
objects (like quasars) with large red shift as consequence of the
waveguiding structure existence between the object and the
observer. This effect exists also for a gravitational lens located
between the object and the observer, but the long gravitational
waveguide may give huge magnification, since the radiation
propagates along the waveguide with functional dependence of the
intensity on the distance which does not decrease as $~\sim
1/R^2~$, characteristic for free propagation. The gravitational
lens, being a compact object, collects much less light by
deflecting the rays to the observer than the gravitational
waveguide structure transporting to the direction of observer all
trapped energy (of course, one needs to take into account losses
for scattering and absorbtion). From that point of view, it is
possible that enormous amount of radiation emitted by quasars is
only seemingly existing. The object may radiate a resonable amount
of energy but the waveguide structure transmits the energy in high
portion to the observer. Similar ideas, related to gravitational
lensing, were discussed in \cite{barn} but, since above mentioned
reasons, the singular lens or even few  aligned strong lenses
cannot produce effect of many orders of magnitude magnification of
brilliancy. The waveguide effect may explain the anomalous high
luminosity observed in  quasars. In fact, quasars are objects at
very high redshift which appear almost as point sources but have
luminosity that are about one hundred times than that of a giant
elliptical galaxy (quasars have luminosity which range between
$10^{38}-10^{41}$ W). For example, PKS 2000-330 has one of the
largest known redshifts $(Z=3.78)$ with a luminosity of $10^{40}$
W. Such a redshift corresponds to a distance of $3700$ Mpc, if it
is assumed that its origin is due to the expansion of the universe
and the Hubble constant is assumed $H=75$km s$^{-1}$ Mpc. This
means that the light left the quasar when the size of the universe
was one--fifth of its present age where no ordinary galaxies
(included the super giant radio--galaxies) are observed. The
quasars, often, have both emission and absorption lines in their
spectra. The emission spectrum is thought to be produced in the
quasar itself; the absorption spectrum, in gas clouds that have
either been ejected from the quasar or just happen to lie along
the same line of sight. The brightness of quasars may vary
rapidly, within a few days or less. Thus, the emitting region can
be no larger than a few light--days, \ie about one hundred
astronomic units. This fact excludes that quasars could be
galaxies (also if most astronomers think that quasars are
extremely active galactic nuclei).

The main question is how to connect this small size with the so
high redshift and luminosity. For example, H.C. Arp discovered
small systems of quasars and galaxies where some of the components
have widely discrepant redshifts \cite{arp}. For this reason,
quasar high redshift could be produced by some unknown process and
not being simply of cosmological origin. This claim is very
controversial. However there is a fairly widely accepted
preliminary model which, in principle,  could unify all the forms
of activities in galaxies (Seyfert, radio, Markarian galaxies and
BL Lac objects). According to this model, most galaxies contain a
compact central  nucleus with mass $10^{7}\div 10^{9}$ M$_{\odot}$
and diameter $< 1 $ pc. For some reason, the nucleus may, some
times, release an amount of energy exceeding the power of all the
rest of the galaxy. If there is only little gas near the nucleus,
this leads to a sort of double radio source. If the nucleus
contains much gas, the energy is directly released as radiation
and one obtains a Seyfert galaxy or, if the luminosity is even
larger, a quasar. In fact, the brightest type 1 Seyfert galaxies
and faintest quasars are not essentially different in luminosity
($\sim 10^{38}$ W) also if the question of redshift has to be
explained (in fact quasar are, apparently, much more distant).
Finally, if there is no gas at all near such an active nucleus,
one gets BL Lac objects. These objects are similar to quasar but
show no emission lines. However the mechanism to release  such a
large amount of energy from active nuclei or quasars is still
unknown. Some people suppose that it is connected to the releasing
of gravitational energy due to the interactions of internal
components of quasars. This mechanism is extremely more efficient
than the releasing of energy during the ordinary nuclear
reactions. The necessary gravitational  energy could be produced,
for example, as consequence of the falling of gas in a very deep
potential well as that connected with a very massive black hole.
Only with this assumption, it is possible to justify a huge
luminosity, a cosmological
 redshift and a small size for the quasars\footnote{However, the question is
extremely controversial and some people do not believe to the
presence of the black hole. An intriguing alternative is that
proposed by Viollier \cite{viollier} who supposes that a heavy
neutrino matter condensation could reproduce the very massive core
of quasars.}.

An alternative explaination could come from waveguiding effects.
As we have discussed, if light travels within a filamentary or a
planar structure, whose Newtonian gravitational potential is
quadratic in the transverse coordinates, the radiation is not
attenuated, moreover the source brightness is stronger than the
brightness of an analogous object located at the same distance
(that is at the same redshift). In other words, if the light of a
quasar undergoes a waveguiding effect, due to some structure along
the path between it and us, the apparent energy released by the
source will be anomalously large, as the object were at a distance
(\ref{eff}). Furthermore, if the approximation $\omega l\ll 1$
does not hold,  the dimensions of resulting image would be
decreased by a factor $\sqrt{1+(\omega l)^{2}}$ while the
brightness would be $(R/R_{eff})^{2}$, larger, then explaining how
it is possible to obtain so large emission by such (apparently)
small  objects. In conclusion, the existence of a waveguiding
effect may prevents to take into consideration exotic mechanism in
order to produce huge amounts of energy (as the existence of a
massive black hole inside a galactic core) and it may justify why
it is possible to observe so distant objects of small geometrical
size.

Another effect concerning the quasars may be directly connected
with multiple images in lensing. The waveguide effect does not
disappear if the axis of ``filament'' or if the guide plane is
bent smoothly in space. As in the case of gravitational lenses, we
can observe  ``twin'' images if part of the radiation comes to the
observer directly from the source, and another part is captured by
the bent waveguide. The ``virtual'' image can then turn out to be
brighter than the ``real'' one (in this case we may deal with
``brothers" rather than ``twins" since parameters like, spectra,
emission periods and chemical compositions are similar but the
brightnesses are different). Furthermore, such a bending in
waveguide could explain large angular separations among the images
of the same object which cannot be explained by the current lens
models (pointlike lens, thin lens and so on).

\begin{figure}[h!]
\begin{center}
\epsfig{file=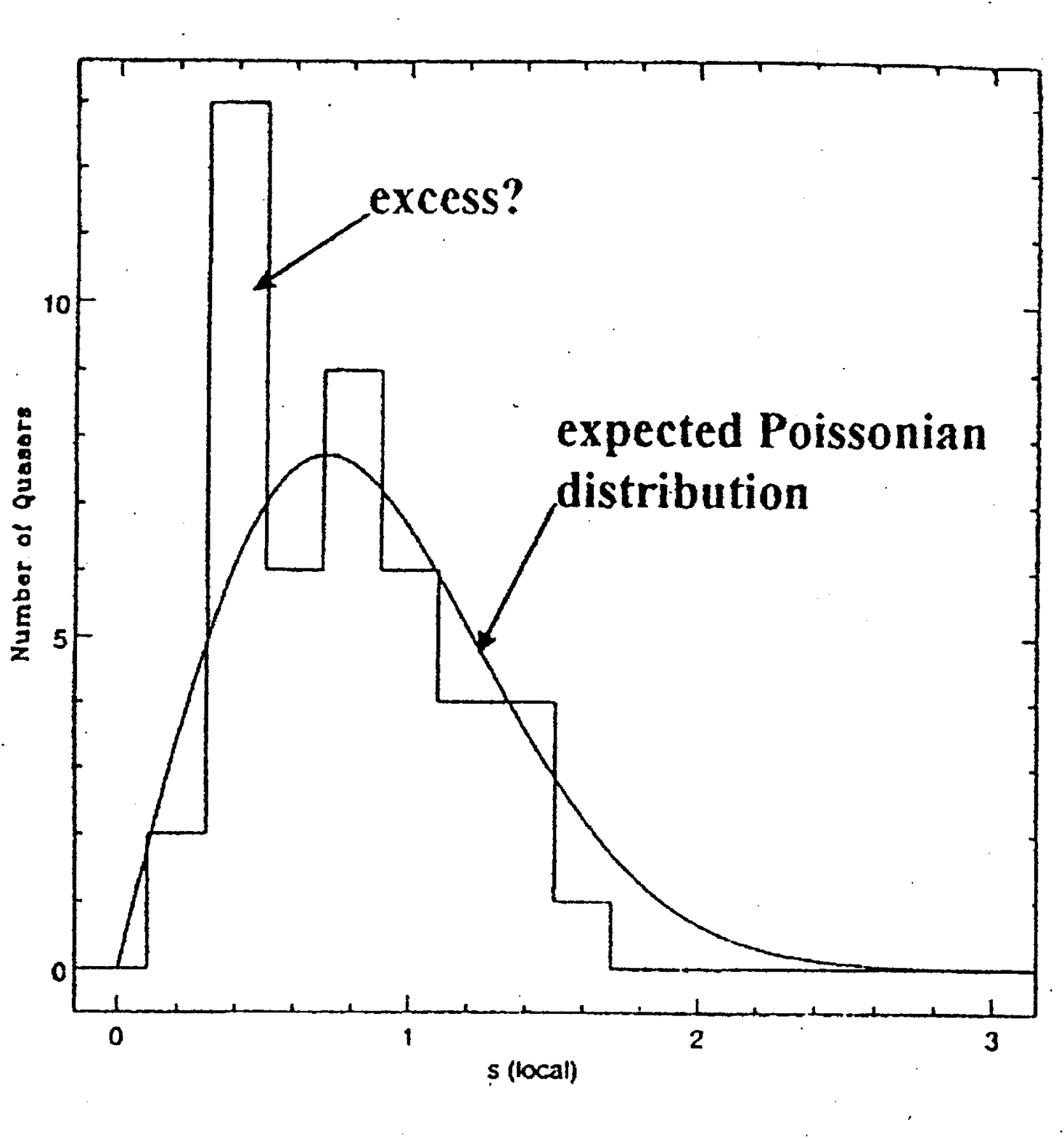,width=8cm,clip=}
\end{center}
\caption{The number of quasars against their local separation. The
excess  could be attributed to some lensing effect. It is clear
the deviation with respect to the expected Poissonian distribution
(CRONA project internal communication).}
\label{fig1}
\end{figure}

\begin{figure}[h!]
\begin{center}
\epsfig{file=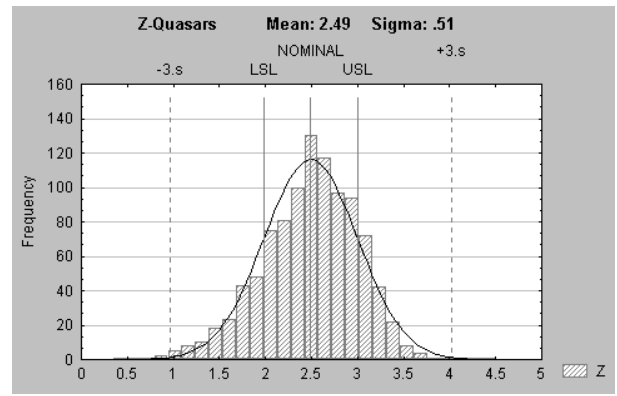,width=8cm,clip=}
\end{center}
\caption{The random  distribution of quasars in redshift $Z$ on a
sphere of radius $\sim 3000$ Mpc. We take into account an
attenuation effect by which we cannot see quasars with a distance
$\geq 1500$ Mpc if the luminosity goes as $r^{-2}$. After this
attenuation, from $10^5$ initial objects, we observe only 980 of
them.} \label{fig2}
\end{figure}

\begin{figure}[h!]
\begin{center}
\epsfig{file=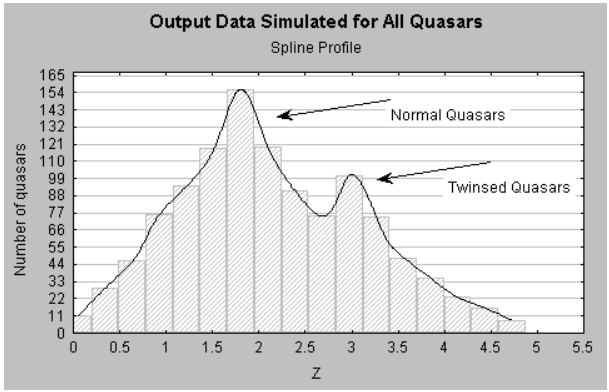,width=8cm,clip=}
\end{center}
\caption{Superposing a random distribution of 200 waveguides (with
sizes $\sim 100$ Mpc of length and $\sim 100$ kpc of thickness) we
get that 137 of the above quasars ($14 \%$) results ``twinsed''.
This simple simulation shows that it is quite easy to get the
double peak in the distribution of quasars by using waveguides.}
\label{fig3}
\end{figure}

This feature could affects also the statistics explaining why the
luminosity distribution of quasars in the sky is not a Poisson
distribution, as expected considering the number of quasars and
their relative distances \cite{bartelmann}. What is observed is a
double peak and the excess could be attributed to some lensing
effect (see Fig.1). Using a simple simulation, which we are going
to explain, it is possible to implement the effect by using a
distribution of waveguides. Let us consider a uniform and
isotropic distribution of quasars ($\sim 10^{5}$) in a 3000 Mpc
sphere, whose peak is at redshift $Z\sim 2.5$. We can take into
consideration also some attenuation effect which selects only
quasars over some brilliancy treshold (Fig. 2). Superposing a
random distribution of filamentary waveguides ($\sim 200$) of
length $\sim 100$ Mpc which give rise to ``twins effect'' as soon
as they interact with quasars (\ie as soon as they catch the
radiation emitted by quasars), the double peak distribution is
reproduced explaining the excess\footnote{By a similar simulation,
it could be possible also to explain the same {\it existence} of
quasars implementing the above mechanism. Given a distribution of
galaxies or protogalaxies (quasars are very old objects if their
redshift has a cosmological origin) which cannot be revealed, the
presence of a distribution of waveguides between them and us
allows their detection.} (see Fig.3).

Now the issue is: are there  cosmic structures which can furnish
workable models for waveguides? Have they to be ``permanent"
structures or may the waveguide effect be accidental (for example
an alignment of galaxies of similar density and structure, due to
cosmic shear and inhomogeneity, may be available as waveguide just
for a limited interval of time \cite{valdes})? In general, both
points of view may be reasonable and here we will outline both of
them. Furthermore we have to consider the problem of the abundance
of such structures: are they  common and everywhere in the
universe or are they peculiar and located in particular regions
(and eras)?

We have to do a first remark on the densities of waveguide
structures which allow observable effects \cite{dodma}.
Considering Eq.(\ref{foc}) and introducing into it the critical
density of the universe $\rho_{c}\sim 10^{-29}$ g/cm$^{3}$ (that
is the value for which the density parameter is $\Omega=1$), we
obtain $L_{foc}\sim 5\times 10^{4}$ Mpc which is an order of
magnitude larger than the observable universe and it is completely
unrealistic. On the contrary, considering a  typical galactic
density as $\rho\sim 10^{-24}$ g/cm$^{3}$, we obtain $L_{foc}\sim
100$ Mpc, which is a typical size of large scale structure (e.g.
the Great Wall  has such dimensions  and also a filament of
galaxies can have such a length \cite{huchra}).

However, an important issue has to be taken into consideration:
the absorption and the scattering of light by the matter inside
the filament or the planar structure increase with density and, at
certain critical value, the waveguide effect can be invalidated
\cite{dodma}. For the smaller frequency of broadcast range (due to
the strong dependence of the absorbtion cross section on the
electromagnetic wavelength) $~\sigma \sim \sigma _T\, (\omega
/\omega _0)^4,$ where Thomson cross-section $~\sigma _T=6\cdot
10^{-25}~\mbox {cm}^2~$ and the characteristic atomic frequency is
$~\omega _0\sim 10^{16}~\mbox {s}^{-1},$ the ratio $~\omega
/\omega _0\ll 1\,,$ and the absorbtion is small. It means that the
absorbtion length $~L_a=m_p/\rho \,\sigma \,,$ (where the mass of
proton $~m_p~$ is approximately equal to the hydrogen atom mass)
is larger than the focusing length $~L_a<L_{foc}$ for the
electomagnetic waves of broadcast range. Thus, the magnification
of electromagnetic waves may be not masked by essential energy
losses due to light absorbtion and scattering processes. However,
no restrictions exist practically  if the radio band
 and a thickness of the structure $r>10^{14}$cm are considered.

In such a case, the relative density change between the background
and the structure density is valid till $\delta \rho/\rho\leq 1$ .
This means that we have to stay in a linear perturbation density
regime.

By such hypotheses, practically all the observed large scale
structures like filaments, walls, bubbles and clusters of galaxies
can result as candidates for waveguiding effect if the
restrictions on density, potential and waveband are respected (in
optical band, such  phenomena are possible but the density has to
be chosen with some care).

However, it is well-known that lensing effects related to large
scale structures, corresponding to density contrasts equal or
smaller than 1, do not give strong lensing effects (in these
situations, we are dealing with ''weak lensing" phenomena). In our
case, the effect is different since the light is ''trapped" and
''guided" inside the structures. Differently, in ordinary lensing
light travels outside the structures. In addition, taking into
account fractal models for large scale structures, one could have
strong gravitational lensing effects thanks to the self-similarity
properties of the models
 (for an exhaustive discussion about the topic, see for example
\cite{labini}). In any case, the waveguide effect could be also
interesting  in a fractal model context: in fact  there is no
contradiction between them.

Also primordial structures (produced in inflationary phase
transition and surviving later), like cosmic string, could furnish
waveguides. In fact, in weak energy limit approximation, such
objects are internally described by the Poisson equation
$\nabla^{2} \Phi=\rho_{0}$ and externally by $\nabla^{2}\Phi=0$
and, furthermore, they act as gravitational lenses after the
formation of the quasar \cite{vilenkin},\cite{gott}. It is easy to
recover an internal potential of the form $\Phi\sim r^{2}$ and,
considering the dynamical evolution after the decoupling, lengths
in the required ranges for waveguide (e.g. $\sim 100$ Mpc). The
main problem is due to the fact that also after the evolution to
macroscopic sizes, strings remain ``wires" without becoming
cylinders, that is their thickness remains well below $r\sim
10^{14}$ cm, the minimal value required to get observable effects.
However, we have not considered the scaling solutions (see for
example \cite{kibble}) from which such wires could evolve in
cylindrical structures (with transverse sizes non trivial with
respect to the background).

Other two interesting features are connected with cosmic strings:
the first is that their motion with respect to the background
produces wakes and filaments which, later, are able to  evolve in
large scale structures systems of galaxies \cite{vachaspati}. For
example, at decoupling $(Z\sim 1000)$, a string can produce a
wake, which consists in a planar structure, with side $\Delta
r\sim 1$ Mpc and constant surface density $\sigma_{0}\sim 3\times
10^{11}$M$_{\odot}$Mpc$^{-2}$. Such a feature is interesting for
large scale structure formation and can yield a planar waveguide
with today observable effects. The second fact is that
inflationary phase transition can produce a large amount of cosmic
strings which, evolving, can give rise to a string network
pervading all the universe \cite{peebles},\cite{vilenkin}. In such
a case, if they evolve in cylindrical or planar structure, we may
expect large probabilities to detect waveguiding effects.

Concerning the second point of view (that is the existence of
temporary waveguiding effects), it could be related to the
observation of objects possessing anomalously large (compared with
their neighbours) angular motion velocities (an analysis in this
sense could come out in mapping galaxies with respect to their
redshift and proper velocities, see for example \cite{davis}).
Such a phenomenon could mean that one observes not the object
itself, but its image transmitted through the moving gravitational
waveguide. The waveguide itself could change its form or it could
be due to temporary alignments of lens galaxies. In this case, the
image of the object could move with essentially different angular
velocity than that of the observable neighbour objects whose light
reaches the observer directly (not throught the waveguide). The
discovery of long distanced objects with anomalous velocity (and
brightness) could support the hypothesis of gravitational
waveguide effect, while the evolution of the waveguide, its
destruction or change of the axis direction (from the orientation
to the Earth) could produce the effect of the disappearence (or
the appearence) of the observed object. For this analysis, it is
crucial to consider long period astronomical observations and deep
pencil beam surveys of galaxies and quasars.

\section{Conclusions}
\sequ{0}

In this paper, we have discussed the possible existence of
gravitational waveguide effects in the universe and constructed a
radiation propagation model to realize them. As in the case of
gravitational lensing, several phenomena and cosmic structures
could confirm their existence, starting from primordial object
like cosmic strings to temporary alignement of evolved late--type
galaxies. Furthermore, due to the wavelength considered, they
could give observable effects in optic, radio or microwave bands
or, alternatively, considering the propagation of other weak
interacting particles as the neutrinos. The experimental
feasibility for the detection could have serious troubles due to
the need of long period observations or due to the discrimination
among data coming from objects which have undergone waveguide
effects and objects which not.

In any case, if such a hypothesis will be confirmed in some of the
above quoted senses, we shall need a profound revision of our
conceptions of large scale structure  and matter distribution.

Finally we want to stress that our treatment does not concern only
electromagnetic radiation: actually a waveguide effect could be
observed
 also for streams of neutrinos \cite{geriov}, gravitational waves
\cite{bimonte},  or other particles which gravitationally interact
with the filament (or the plane), in this sense it could result
useful also in other fields of astrophysics and fundamental
physics.

\vfill

\small
\begin{thebibliography}{99}

\bibitem{peebles}
P.J.E. Peebles  {\it Principles of
Physical Cosmology} (Princeton Univ. Press.,
Princeton 1993)
\bibitem{borgeest}
U. Borgeest \aa {\bf 128} 162 (1983)
\bibitem{frieman}
J.A. Frieman, D.D. Harari, and G.C. Surpi
\pr {\bf D 50} 4895 (1994)
\bibitem{fort}
B. Fort, and Y. Mellier \aa {\it Rev.}  {\bf 5} 239 (1994)
\bibitem{broadhurst}
T.J. Broadhurst, A.N. Taylor, and J.A. Peakock
\apj {\bf 438} 49 (1995)
\bibitem{mollerach}
S. Mollerach and E. Roulet, \prep {\bf 279}, n.2 (1997).
\bibitem{paczynski}
B. Paczynski {\it Gravitational Lenses}
Lecture Notes in Physics {\bf 406}, p. 163, Springer--Verlag, Berlin (1992)
\bibitem{kaiser}
R. Kaiser {\it Gravitational Lenses}
Lecture Notes in Physics {\bf 404}, p. 143, Springer--Verlag, Berlin (1992)
\bibitem{bartelmann}
M. Bartelmann and  P. Schneider \aa {\bf 268}, 1 (1993),\\
M. Bartemann and P. Schneider \aa {\bf 284}, 1 (1994).
\bibitem{kormann}
R. Kormann, P. Schneider, and M. Bartelmann
\aa {\bf 284}, 285 (1994)
\bibitem{walsh}
D. Walsh, R.F. Carswell,  and R.J. Weymann {\it Nat.} {\bf 279}, 381 (1979)
\bibitem{blioh}
P.V. Blioh and A.A. Minakov {\it Gravitational Lenses}
Kiev, Naukova Dumka (1989) (in Russian)
\bibitem{schneider}
P. Schneider {\it Cosmological Applications of Gravitational Lensing},
Lecture Notes in Physics, eds. E. Martinez--Gonzales, J.L. Sanz,
Springer Verlag, Berlin (1996), Astro--Ph/9512047 (1995).
\bibitem{ehlers}
P.Schneider, J. Ehlers, and E.E. Falco {\it Gravitational Lenses}
Springer--Verlag, Berlin (1992).
\bibitem{lifsits}
E. M. Lifsits, L.P. Pitaevskij {\it Elettrodinamica dei mezzi continui},
Ed. Riuniti, Roma (1986).
\bibitem{dodma}
V.V. Dodonov and  V.I. Man'ko, {\it Gravitational waveguide},
Preprint of the  Lebedev Physical Institute Proceedings, No. 255
(Moscow, 1988); {\it J. Soviet Laser Research} (Plenum Press), {\bf 10},
240 (1989); {\it Invariants and Evolution of Nonstationary Quantum Systems}
Proceedings of the Lebedev Physical Institute, {\bf 183}
 Nova Science Publishers N.Y. (1989).
\bibitem{dodonov}
V.V. Dodonov in:
{\it Proceedings of the First International Sakharov Conference}
Moskow May 21--25 (1992), p. 241
Edt. L. V. Keldysh, V. Ya. Fainberg, Nova Science Publishers N.Y. (1993)
\bibitem{dodma1}
V.V. Dodonov, O.V. Man'ko, and V.I. Man'ko, in
{\it Sqeezed and Correlated States of Quantum Systems}
Proceedings of the Lebedev Physical Institute, {\bf 205} p. 217
 Nova Science Publishers N.Y. (1993).
\bibitem{turner}
E. L. Turner, D. P. Schneider, B. F. Burke, J. N. Hewitt,
G. L. Langston, J. E. Gunn, C. R. Lawrence, and M. Schmidt, {\sl Nature},
{\bf 321} 142 (1986).
\bibitem{labini}
F. Sylos Labini, M. Montuori, L. Pietronero, \prep {\bf 293}, 61
(1998).
\bibitem{vilenkin}
A. Vilenkin, \apj {\bf 282}, L51 (1984);\\
A. Vilenkin, \prep {\bf 121}, 263 (1985).
\bibitem{gott}
J.R. Gott III, \apj {\bf 288}, 422 (1985).
\bibitem{fock}
A. M. Leontovich and V. A. Fock, {\sl Zh. Eksp. Teor.
Fiz.}, {\bf 16}, 557 (1946).
\bibitem{manko}
V. I. Man'ko, in ~{\it Lee Methods in Optics}, Lecture
Notes in Physics, Eds. S. Mondragon and K.-B. Wolf, {\bf 250} 193 (1986).
\bibitem{kolb}
E.W. Kolb, M.S. Turner {\it The Early Universe}
(Addison--Wesley Pub. Co., Menlo Park 1990)
\bibitem{binney}
J. Binney,  S. Tremaine  {\it Galactic Dynamics} (Princeton
Univ. Press.: Princeton 1987)
\bibitem{mihalas}
D. Mihalas, J.J. Binney {\it Galactic Astronomy}
2nd. ed. San Francisco, Freeman (1981)
\bibitem{refsdal}
S. Refsdal, J. Surdej {\it Rep. Prog. Phys.} {\bf 57}, 117 (1994).
\bibitem{landau}
L. D. Landau and E. M. Lifsits {\it The Classical Theory of Fields},
Pergamon Press, N.Y. (1975).
\bibitem{skrotsky}
G.V. Skrotsky, {\it Doklady} AN SSSR, {\bf 114}, 73 (1957).
\bibitem{plebansky}
J. Plebansky, \pr, {\bf 118}, 1396 (1960).
\bibitem{arp}
H.C. Arp {\it Quasars, Redshifts and Controversies} Berkeley: Interstellar
Media, $\$ 7$ (1987).
\bibitem{viollier}
R.D. Viollier {\it Prog. Part. Nucl. Phys.} {\bf 32}, 51 (1994).
\bibitem{valdes}
F. Valdes, J.A. Tyson, J.F. Jarvis \apj {\bf 271}, 431 (1983).
\bibitem{huchra}
V. de Lapparent, M.J. Geller, J.P. Huchra \apj {\bf 302}, L1 (1986).\\
M.J. Geller, J.P. Huchra {\it Science} {\bf 246}, 897 (1989).
\bibitem{kibble}
T.W.B. Kibble {\it J. Phys.} {\bf A9}, 1387 (1976).
\bibitem{vachaspati}
T. Vachaspati, A. Vilenkin, \prl {\bf 67}, 1057 (1991).
\bibitem{seitz}
S. Seitz, P. Schneider, J. Ehlers, \cqg {\bf 11}, 23 (1994).
\bibitem{davis}
M. Davis, J. Huchra, D.W. Latham, J. Tonry \apj {\bf 253}, 423 (1982)
\bibitem{barn}
J. M. Barnothy, {\sl Astron. J.}, {\bf 70}, 666 (1965).
\bibitem{geriov}
S. Capozziello and G. Iovane, {\it Proc. of II Int. Conf. on Dark Matter in
Astro and Particle Physics}, Heidelberg July 1998, Ed.
Klapdor--Kleingrothaus,
IOP Publishing Bristol (1998).
\bibitem{bimonte}
G. Bimonte, S. Capozziello,  V. Man'ko, G. Marmo,  \pr {\bf 58 D}
n.10, 104009 (1998).

\end{thebibliography}
\end{document}